\documentstyle[preprint,aps]{revtex}

\begin{document}
\draft

\title{Quasi-equilibria in one-dimensional self-gravitating many body systems}
\author{Toshio Tsuchiya\cite{Email}}
\address{Department of Physics, Kyoto University, Kyoto, 606-01, Japan}
\author{Tetsuro Konishi}
\address{Department of Physics, Nagoya University, Nagoya, 464-01, Japan}
\author{Naoteru Gouda}
\address{Department of Earth and Space Science, Osaka University,
Toyonaka, 560, Japan}
\date{\today}
\maketitle

\begin{abstract}
The microscopic dynamics of one-dimensional self-gravitating many-body
systems is studied. We examine two courses of the evolution which has
the isothermal and stationary water-bag distribution as initial
conditions. We investigate the evolution of the systems toward thermal
equilibrium.  It is found that when the number of degrees of freedom of
the system is increased, the water-bag distribution becomes a
quasi-equilibrium, and also the stochasticity of the system reduces.
This results suggest that the phase space of the system is effectively
not ergodic and the system with large degreees of freedom approaches to
the near-integrable one.
\end{abstract}

\narrowtext

\section{Introduction}
\label{sec:intro}
A self-gravitating many body system, which contains many particles
interacting with mutual gravity, is an idealized model of wide classes
of astronomical objects, such as globular clusters, elliptical galaxies,
and clusters of galaxy. Dynamics of the system has two characteristic
aspects. One is the {\em microscopic dynamics}, which is concerned with
the motions of the individual particles. The other is the {\em
macroscopic dynamics}, which deals with the averaged quantities.
Evolution of the real system is determined exactly by the microscopic
dynamics. However, the number of the particles contained in the system
is so large, e.g. $10^5$ for globular clusters and $10^{11}$ for
elliptical galaxies, that we can treat only macroscopic quantities by
the statistical method, in practice.

In many cases which are successfully treated by statistical mechanics,
which include gas and liquid systems, the particles interact
with close neighbours, then the local equilibrium is determined by the
state of the neighbours. Thus if the macroscopic quantities are defined
by averaging over the scale much larger than the range of the force,
their evolution is independent of the microscopic dynamics and is
governed only by themselves . In these systems the statistical mechanics
is easily applicable.

On the other hand, in the self-gravitating systems, the motions of the
individual particles are governed by the summation of the forces from
all the other, because the gravity is long range force. Thus the
macroscopic dynamics is not decoupled with the microscopic dynamics.
Therefore we must study how the microscopic dynamics influences the
evolution of the macroscopic quantities in the self-gravitating systems.

The microscopic dynamics is described as a trajectory in the
$\Gamma$-space ($N$-body phase space). If the system has some integrals
of motion, the trajectory is confined in the subspace which conserves
the integrals. The system is defined as {\em ergodic\/} if the time average
of any dynamical quantities are equal to the spatial average
over the subspace. This subspace is referred to as the ergodic region. In
this case the time average does not depend on the initial condition of the
system but only on the integrals, and thus it is time independent. This
gives the statistical or micro-canonical equilibrium. At the
equilibrium, energy is equally distributed to all degrees of freedom,
which is called {\em equipartition}.

The mixing system, which is included in the ergodic system, has the
property that a small but finite part of the phase space volume spreads
over whole ergodic region by means of coarse-graining. This property is
closely related to relaxation of the system, because information of the
initial state is lost and the micro-canonical equilibrium is realised.
Any time-correlation disappears after the trajectory is mixed over the
ergodic region [for more detailed reviews, see, e.g.
\cite{arn68,lic83}]. We refer to the time that mixing is realized as the
{\em microscopic relaxation time}.

Though ergodicity is the fundamental basis of statistical mechanics, not
all systems are ergodic. For example, the phase space of a
near-integrable Hamiltonian system contains both stochastic region and
regular region (tori). A trajectory initially located on a torus stays
on it forever. This system is not ergodic over the subspace where the
total energy of the system is constant. Furthermore, it is shown that
there exist {\em stagnant layers\/} around tori and a trajectory in the
stagnant layer stays there for a long time \cite{aiz89b}. In this
system, the trajectory in the stagnant layer moves in the stochastic
region after a long interval of time, and then they could enter another
stagnant layer. Thus, in the system, transient stationary states can
emerge in some era. We refer to them as {\em quasi-equilibria}.

Lynden-Bell applied the concept of ergodicity to the evolution of
elliptical galaxies, and derived the unique equilibrium state
\cite{lyn67}.  However, observations of elliptical galaxies and
numerical simulations show disagreement with the Lynden-Bell
distribution \cite{dez91,fun92a,yam92}. In particular, elliptical
galaxies are believed to be triaxial in the shape and anisotropic in the
velocity dispersion. This stationary state seems to suggest the
existence of additional integrals, which conserves the anisotropy,
though it is unclear whether the state is a stationary state induced by
the additional integrals or only a
transient state approaching to the statistical equilibrium. These facts
suggest that elliptical galaxies are not ergodic while it is generally
believed that the self-gravitating many body systems are chaotic and so
ergodic. So it is very important and interesting to study ergodicity of
the self-gravitating many body systems in order to analyse the present
dynamical structures of elliptical galaxies, and moreover to examine
applicability of usual statistical mechanics to the self-gravitating
many-body systems.

To study their ergodicity, we employ one-dimensional system. Because in
one-dimension the phase space is compact, which makes the system
tractable in considering ergodicity.  Another reason is that the force
law is very simple and thus the evolution of the system can be followed
numerically with a good accuracy by using exact code\cite{hoh67b}.
Though the force law in one-dimensional system is different from that in
the three-dimensional system, we can study the properties induced by
long range forces even in the one-dimensional systems.

Hohl\cite{hoh67a,hoh67b} firstly suggested that the one-dimensional
system consisting of $N$ identical plane-parallel sheets should relax on
the order of $N^2t_c$, where $t_c$ is the characteristic time which is
approximately the time for a member to traverse the system. However,
Wright, Miller, and Stein\cite{wri82} asserted that some initial states
does not approach to micro-canonical distribution after $2N^2t_c$. Thus
the Hole's conjecture was questioned. In succeeding paper
\cite{luw84,sev84a,luw85,rei87,rei88,mill90,rei91}, it was shown that
the evolution of the system greatly depends on the initial condition.
Some initial state appears to relax on the time scale of $Nt_c$, which
is much shorter than Hohl's prediction \cite{luw84,sev84a,rei92}. Then
the complicated features of relaxation in the self-gravitating systems
were recognized.

Severne and Luwel \cite{sev86} suggested that there are three phase in
relaxation. If the initial state is far from equilibrium, violent
oscillation of mean field gives rise to the violent
relaxation\cite{lyn67} for the first several oscillation. After the
system almost virialised, remaining small fluctuation of gravitational
field causes the change of the individual particle energies. They called
this era as the collisionless mixing phase. In astrophysics the
evolution in the violent relaxation and collisionless mixing phase are
often referred to dynamical evolution. After that, the collisional
relaxation phase takes place, in which the particle interactions tend to
drive the system towards the microscopic thermal equilibrium. This is
the thermal evolution. Luwel and Severne\cite{luw85} showed that the
collisionless mixing occurs in the stationary water-bag distribution.
The distribution is one of the stationary solutions of the collisionless
Boltzmann-Poisson equations.  Thus it does not change its macroscopic
distribution in collisionless phase. Luwel and Severne did not study the
evolution in the collisional phase. If the system is ergodic, it is
expected that the stationary water-bag distribution is transformed into
the microcanonical distribution after the collisional relaxation becomes
effective.

While the above studies are concerned with the evolution of the
macroscopic quantities, the microscopic dynamics, especially the
ergodicity of the system also has been studied by some authors. For
small $N$ $(N\leq10)$ it is shown that the systems are ``ergodic'' for
$N>5$\cite{fro75,ben79,wri84}. For $11\leq N \leq20$, Reidl and
Miller\cite{rei92} found that two different periodic orbits have the
different maximum Lyapunov exponents.  From this fact they asserted that
the trajectories of the two orbits in the phase space covers different
regions, which are separated from each other.

However, this conclusion is doubtful because the measure of the
trajectory of the periodic orbit in the phase space is negligible then
the study gave no information of the whole phase space dynamics.
Further, the numbers of the sheets are too small to speculate the
macroscopic relaxation in larger $N$. Thus we performed numerical
simulations for $N=32$, 128, 512, and examine the dependence of the
dynamics on the number of the sheets (i.e., the degrees of freedom of
the system).

In order to get more information of microscopic dynamics, we have
investigated the time correlation of the fluctuation of the individual
particle energies, the equipartition of the individual particle energies
and the convergence of the maximum Lyapunov exponent. In general, the
motion in an ergodic region shows fast decay of correlation. On the
other hand, in the near integrable Hamilton system an essential feature
of the diffusion process (including the Arnold diffusion) is the
appearance of long time tails of the correlation or the enhancement of
the diffusion mode with the zero frequency, e.g., the power spectrum
density (PSD) function $S(f)$ satisfies,
\begin{equation}
S(f)\sim f^{-\nu}, \quad (f\ll1)
\end{equation}
where $f$ stands for the frequency and $\nu$ a positive constant
($2>\nu>1$)\cite{aiz89b,zac86}. Therefore the time correlation or the
PSD is a good tool to understand the ergodicity of the phase space,
besides the equipartition of the energies and the convergence of the
maximum Lyapunov exponents.

In section \ref{sec:model} we describe the model and the initial
conditions. Three quantities which we employ to analyse the system are
explained in section \ref{sec:analyses} and results of the numerical
simulations are given in section \ref{sec:result}. We devote section
\ref{sec:conclusion} to the conclusions and discussions.

\section{Description of the model}
\label{sec:model}
The one-dimensional self-gravitating system consists of $N$ identical
mass sheets, each of uniform mass density and infinite in extent in the
$(y,z)$-plane. We call the sheet as particle in this paper. The
particles are free to move along $x$-axis and accelerate as a result of
their mutual gravitational attraction. The Hamiltonian of the system is
given by
\begin{equation}
H=\frac{m}{2}\sum_{i=1}^N v_i^2  + (2\pi Gm^2)\sum_{i<j} |x_j-x_i|,
\end{equation}
where $m$, $v_i$, and $x_i$ are the mass (surface density), velocity,
and position of the $i$th particle, respectively. Since the
gravitational field is uniform, the individual particles moves
parabolically, until they intersect with the neighbours. Thus the
evolution of the system can be followed by connecting the parabolic
motions. When an encounter occurs between two particles, they pass
freely through each other.

Our code is the application of that for one-dimensional sheet
plasma\cite{mit92} to the self-gravitating systems, which is very similar
to the ``new exact code'' referred in ref\cite{sev86}, but the exchange
of particles at an encounter is arranged by the heap sort algorithm.
During the integration, time is measured in the unit
\begin{equation}
t_c=(4\pi G\rho_{\text{av}})^{-1/2},
\end{equation}
where $\rho_{\text{av}}$ is the mass divided by the width of the
distribution at the initial time.

All calculation were performed in double precision (16 significant
figures) on DEC station 3000AXP and a SONY NEWS 5000. The total energy
was conserved to better than one part in $10^{13}$.

We examined two courses of evolution, IT and WB, which begin with the
isothermal and water-bag distributions as their initial condition,
respectively. In the water-bag distribution, all particles are randomly
distributed in a rectangle region in the $\mu$-space to form a uniform
averaged phase density (Fig.\ref{fig:WB}). This is not a exact
stationary solution of the collisionless Boltzmann-Poisson equations
because the shape is a rectangle, but the difference is very small and
then after several oscillations it is transformed into the stationary
water-bag configuration. The isothermal distribution is given by
\begin{eqnarray}
\theta(\eta)&=&\pi^{-1/2} \exp(-\eta^2) \quad \hbox{(velocity)}, \\
\rho(\xi)&=&\frac{1}{2}{\rm sech}^2\xi \quad \hbox{(position)},
\end{eqnarray}
where
\begin{equation}
\eta=(v/2)(3M/E)^{1/2},
\end{equation}
and
\begin{equation}
\xi=(3\pi GM^2/2E)x,
\end{equation}
$M$ and $E$ represent the total system mass and total energy,
respectively(Fig. \ref{fig:IT}). It is a stationary solution of the
collisionless Boltzmann-Poisson equations. In all our simulations, the
total mass and energy set to 1 and 1/4, respectively. The typical period
of oscillation of a particle is $2\pi t_c$.

We choose the initial distributions which are dynamically stable,
because our studies are concentrated on the thermal evolution
(collisional relaxation) of the system. If the system is ergodic, IT and
WB should coincide after the relaxation time. Therefore we compare the
behaviours of IT and WB for a long time.

\section{Analyses}
\label{sec:analyses}
\subsection{Equipartition}
The specific energy (energy per unit mass) $\varepsilon_i(t)$ of $i$th
particle is given by
\begin{equation}
\varepsilon_i(t)=\frac{1}{2}v_i^2(t) + 2\pi Gm \sum_{j=1}^N|x_j(t)-x_i(t)|.
\end{equation}
If the evolution of the system is ergodic in the $\Gamma$-space, the
long time average of the specific energy takes a unique value for all
$i$, i.e.
\begin{equation}
\overline{\varepsilon_i}\equiv \lim_{T \rightarrow \infty} \frac{1}{T}
\int_0^T \varepsilon_i(t) dt = \varepsilon_0 \equiv 5E/3.
\end{equation}
The degree of the deviation from the equipartition is measured by
the quantity,
\begin{equation}
\Delta(t)\equiv \varepsilon_0^{-1} \sqrt{\frac{1}{N} \sum_{i=1}^N
(\overline{\varepsilon_i}(t) - \varepsilon_0)^2},
\end{equation}
where $\overline{\varepsilon_i}(t)$ is the averaged value until $t$. In
the numerical scheme, $\varepsilon_i(t)$s are sampled at every $\Delta
t=0.78125t_c$, and the average is defined simply by the summation of the
samples divided by the number of the samples.

If the system is mixing, we can estimate the way of the evolution of
$\Delta(t)$. In the system, time correlation disappears in a finite time
(the relaxation time). In this time, a trajectory in the $\Gamma$-space
visits almost every point in the ergodic region, thus the $\Delta(t)$
almost vanishes but gives a small fluctuation. Since the correlation
vanishes after the relaxation time, $\Delta(t)$ decreases as $t^{-1/2}$
for the time longer than the relaxation time, according to the central
limit theorem.  Therefore the evolution of $\Delta(t)$ is one of good
tests of mixing property of the system.

The initial value of the deviation from the equipartition, $\Delta(0)$,
depends on the initial distribution. Figure \ref{fig:edist} shows the
cumulative distribution of the specific energy,
$\hat{\nu}(\varepsilon)$, which is defined by
\begin{equation}
\hat{\nu}(\varepsilon)\equiv (1/N) N(\varepsilon_i \leq \varepsilon),
\end{equation}
where $N(\varepsilon_i \leq \varepsilon)$ is the number of particles with
$\varepsilon_i \leq \varepsilon$. The isothermal and water-bag
distribution are represented by the solid and dashed lines,
respectively. Also the energy at the equipartition, $\varepsilon_0$, is
shown by dotted line. Since the isothermal distribution spreads wider in
the energy space than the water-bag, $\Delta(0)$ of the isothermal
distribution is larger than that of the water-bag distribution.

\subsection{Power spectrum density}
The power spectrum density (PSD) $S(f)$ is given by
\begin{equation}
S(f)\equiv (1/N) \sum_{i=1}^N |C_i(f)|^2,
\end{equation}
where $C_i(f)$ is Fourier transform of $\varepsilon_i(t)$, i.e.
\begin{equation}
\varepsilon_i(t)=\int C_i(f) {\rm e}^{-2\pi ift} df.
\end{equation}
The PSD, on the other hand, is the Fourier transform of the
autocorrelation function, thus the long time correlation gives rise to a
peak at small $f$. In order to obtain PSD numerically, a sequence of
``locally averaged'' energy $\{ \langle \varepsilon \rangle (t_1=T_0/n),\,
\langle\varepsilon\rangle(t_2=2T_0/n),\,  \ldots,\,
\langle\varepsilon\rangle(t_n=T_0)\}$ are sampled, where
\begin{equation}
\langle\varepsilon\rangle{}_i (t_j)=\frac{1}{t_1}\int_{t_{j-1}}^{t_j}
\varepsilon_i(t) dt.
\label{eq:locave}
\end{equation}
The maximum integration time $T_0$ determines the minimum frequency
$f_{\text{min}}=1/T_0$, and the interval of sampling determines the
maximum frequency $f_{\text{max}}$.  This local averaging suppresses the
higher frequency modes than $f_{\text{max}}$, thus it prevents the
``aliasing'', which means that the higher frequency modes fall into the
interval of $f\leq f_{\text{max}}$ \cite{pre88}. The procedure of the
local averaging, eq. (\ref{eq:locave}), is the same as that described in
the previous section. We integrated the motion to $T_0=10^7t_c$ for
$N=$32 and 128, hence $f_{\text{min}}=10^{-7}t_c^{-1}$. The
number of samples $n$ is limited to 256 due to the computer's ability,
hence $f_{\text{max}}=2.56\times 10^{-5}t_c^{-1}$.  In order to obtain
higher $f$, we gathered $n$ samples with shorter time interval,
$\{\langle \varepsilon_i \rangle (T_0/10n), \langle
\varepsilon_i \rangle (T_0/10n), \langle \varepsilon_i \rangle
(2T_0/10n), \cdots, \langle \varepsilon_i \rangle (T_0/10)\}$, which
covers the range of the frequency from $10^{-6}t_c^{-1}$ to
$2.56\times10^{-4}t_c^{-1}$. If the amplitude of oscillation with the
frequency
$10^{-6}t_c^{-1}\leq f \leq 2.56\times10^{-5}t_c^{-1}$ for
$t\geq10^6t_c$ is the same as during $t\leq10^6t_c$, two $S(f)$ merge
into one curve. In this way we extended the range of the frequency
beyond $10^{-2}t_c^{-1}$. The system with $N=128$ is the exception that
the values of $S(f)$ of $T_0=10^7t_c$ and $T_0\leq10^6t_c$, for $10^6
t_c^{-1}\leq f \leq 2.56\times10^{-5}t_c^{-1}$ are different(Fig.
\ref{fig:bi7b}). This is discussed in the next section.


Since the PSD is the Fourier transform of the autocorrelation function
\begin{equation}
\xi_{\varepsilon}(t)\equiv \int \varepsilon(t+t')\varepsilon(t') dt',
\end{equation}
we can find the time scale on which the correlation decays. For example,
the Brownian motion, $u(t)$, subjected by the Langevin equation,
\begin{equation}
\frac{du}{dt}+\gamma u = R(t),
\end{equation}
where $\gamma$ is the friction coefficient and $R(t)$ is a random force
(white noise), gives the PSD of the Lorentz distribution,
\begin{equation}
S(f) \propto \frac{1}{f^2+\gamma^2}.
\end{equation}

For $f\gg\gamma$, $S(f)\propto f^{-2}$, which describes the short time
scale nature of the system. In this phase, the random force $R(t)$ is
dominated and the dispersion $\langle(u(t)-u(0))^2\rangle$ increases
proportional to $t$. In other words, it is the diffusion phase from its
initial value. On the other hand, for $f\ll\gamma$, $S(f)$ is constant.
Thus any correlation disappears in the time scale $\gamma^{-1}$, and
then the fluctuation becomes like a thermal noise. Since this time scale
is closely related to the relaxation time, the PSD gives us a
complemental information about the mixing property of the system.

We should mention more about the PSD with a peak or divergence as $f$
goes to zero, e.g. $S(f)\propto f^{-1}$, which means the existence of a
long time correlation. Such a PSD is often observed in a near-integrable
system. In the system, the phase space is divided into stochastic region
and regular region (tori). It is well-known as the Arnold diffusion that
it takes a long time for a trajectory to travel across the web of the
tori. This slow diffusion gives rise to a long time
correlation\cite{aiz89b}.

\subsection{Lyapunov exponent}
The evolution of the system can be described a trajectory in the
$\Gamma$-space. The Lyapunov exponents of a given trajectory
characterized the mean exponential rate of divergence of trajectories
surrounding it [a general discussion of the Lyapunov exponents is
provided by Lichtenberg and Lieberman\cite{lic83}].

The maximum Lyapunov exponent is defined as
\begin{equation}
\lambda=\lim_{\stackrel{d(0)\rightarrow 0}{t\rightarrow \infty}} \frac{1}{t}
\ln\frac{d(t)}{d(0)},
\end{equation}
where $d(t)$ and $d(0)$ are the separations in the $\Gamma$-space
between two nearby orbits at times $t$ and 0, respectively. The
numerical procedure follows mostly Shimada and Nagashima\cite{shi79},
where $d(t)$ is determined by the linearized equations of motion:
\begin{eqnarray}
\Delta \dot{x}_i &=& \Delta v_i, \\
\Delta \dot{v}_i &=& -4\pi Gm \sum_{j=1}^N (\Delta x_i-\Delta x_j)
\delta(x_i-x_j), \\
d&\equiv& \sqrt{\sum_{j=1}^N(\Delta x_i^2 + \Delta v_i^2)},
\end{eqnarray}
where $\Delta x_i$ and $\Delta v_i$ are the first order deviation of the
position and velocity. Since $d(t)$ diverges exponentially, when the
separation $d$ becomes $10^5$ times the initial value, the separation is
rescaled to $d(0)$.

The Lyapunov exponent is originally defined by the limit of infinite
$t$, but here we use ``time-dependent Lyapunov exponent'' defined by
\begin{equation}
\lambda(t)=\frac{1}{t} \ln\frac{d(t)}{d(0)}.
\end{equation}
If the system is mixing, the time average becomes roughly the same as
the space average over ergodic region, in the relaxation time. Thus the
convergence of $\lambda(t)$ gives another measure of relaxation.

\section{Results}
\label{sec:result}
\subsection{$N$=32}
Figure \ref{fig:bi5a} shows the time evolution of the deviation from the
equipartition, $\Delta(t)$ (bold lines) and the Lyapunov exponents,
$\lambda(t)$ (light lines). The PSD, $S(f)$, are shown in Fig.
\ref{fig:bi5b}.  The curves of IT and WB are represented by solid and
dashed lines, respectively.

The features of the curves are not so simple as we expected in the
previous section. $\Delta(t)$ begins to decrease at
$t\sim3\times10^3t_c$, but does not show monotonic decrease like
$t^{-1/2}$. From $t\sim3\times10^3t_c$, $\Delta(t)$ of IT and WB
coincide, but at $t\sim10^4t_c$, IT begins to increase again and the
difference between IT and WB lasts until $t\sim3\times10^6t_c$. After
that, though the range of $t$ is not so long, $\Delta(t)$ seems to
decrease as $t^{-1/2}$. At this time, $t\sim3\times10^6t_c$, also the
convergence of $\lambda(t)$ becomes quite well. From the PSD
(Fig.~\ref{fig:bi5b}), it is found that $S(f)$ is constant for $f\ll
10^{-5}t_c^{-1}$, which gives the time scale of disappearance of
correlation of $t\sim10^5t_c$. Therefore it is probable that the system
relaxes on this time, $t\sim10^6t_c$.

{}From Fig. \ref{fig:bi5a}, there seems to exist another time scale,
$t\sim10^3t_c$, at which $\Delta(t)$ begins to decrease. For shorter
period, $t \ll 10^3t_c$, the PSD indicates that the variation of
$\varepsilon_i(t)$ is well-approximated by the random walk because
$S(f)\propto f^{-2}$. For $t\gtrsim 3\times10^3t_c$, the trajectory in
the $\Gamma$-space travels farther so that the individual particle
energies vary widely over the energy space. However, the long time
correlation seems to last for several $10^4t_c$, because
$\Delta(t)$ does not decrease monotonically like $t^{-1/2}$ and
convergence of $\lambda(t)$ is not good. This correlation is not swept
away by the diffusion due to the random walk.

{}From these facts, it is speculated that the system has a part of mixing
property with unique ergodic region, but the phase space is not uniform,
and contains many structures, maybe a set of ruined tori, which yield
the long time correlation for $t<10^5t_c$.

\subsection{$N$=128}
Figure \ref{fig:bi7a} and \ref{fig:bi7b} are the same as Fig.
\ref{fig:bi5a} and \ref{fig:bi5b}, but for $N=128$.
The PSD for WB is that $S(f)$ is separated into two part,
$10^{-7}t_c^{-1} \leq f \leq 2.56\times10^{-5}t_c^{-1}$ and
$10^{-6}t_c^{-1} \leq f \leq 2.56\times10^{-2}t_c^{-1}$. The former is
calculated from the sample of $T_0=10^7t_c$ and $n=256$ (see \S
\ref{sec:analyses}B).  The other is calculated from that of
$T_0\leq10^6t_c$, thus this part does not contain the behaviour during
$10^6t_c\leq t$. The difference between the two means that
evolution with a period $10^6t_c$ later than $t=10^6t_c$ is different
from that of $0\leq t \leq 10^6t_c$.  The transition of state at $t\sim
10^6t_c$ is found in Fig.
\ref{fig:bi7a}. $\Delta(t)$ of WB decreases as $t^{-1/2}$ until $t\leq
10^6t_c$ and then suddenly increases to approach to IT. Also
$\lambda(t)$ shows the transition at $t\sim2\times10^6t_c$. After that
$\Delta(t)$ and $\lambda(t)$ of WB show tendency to approach to that of
IT. Therefore the system is expected to relax on the time scale of
$10^7t_c$.

Another interesting feature of this system is the behaviour of WB for
$t<10^6t_c$. $\Delta(t)$ shows a good agreement with $t^{-1/2}$ from
$t\sim 10^2t_c$, and $\lambda(t)$ converges to the local value,
$5.5\times10^{-2}$, while that of IT converges to $6.2\times10^{-2}$.
Furthermore, $S(f)$ calculated from the sample of $T_0\leq10^6t_c$ shows
almost flat spectrum for $f<10^{-3}t_c^{-1}$. These features indicate
that WB relaxes in $t\sim10^3t_c$ to a quasi-equilibrium which lasts
until $t\sim10^6t_c$. Figure \ref{fig:bi7c} shows the cumulative energy
distribution, $\hat{\nu}(\varepsilon)$, of IT at the beginning, $t=0$,
(solid line), WB at the beginning (dashed line), and WB at $t=10^6t_c$
(dotted line). The distribution of WB at $t=10^6t_c$ is very close to
the initial distribution compared with the isothermal distribution.
Therefore the quasi-equilibrium has, in fact, the water-bag distribution.
On the other hand, the difference between IT and WB disappears after the
transition. Figure \ref{fig:bi7d} shows $\hat{\nu}(\varepsilon)$ of IT
(solid line) and WB (dashed line) at $t=10^7t_c$.

The evolution of IT is similar to that of $N=32$. There is a long time
correlation, which is found from the evolution of $\Delta(t)$ and
$\lambda(t)$. Especially for $N=128$, PSD shows clear bend at
$f=2\times10^{-4}t_c^{-1}$, and the power of $f$ for $10^{-6}t_c^{-1} <
f < 10^{-4}t_c^{-1}$ is less steep than $f^{-2}$.

\subsection{$N=512$}
Figure \ref{fig:bi9a} and \ref{fig:bi9b} show the same as Fig.~
\ref{fig:bi5a} and \ref{fig:bi5b} but for $N=512$, respectively.

The PSD indicates that the behaviour in a shorter time scale than
$10^3t_c$ is the same for IT and WB. It is the diffusion process of the
individual particle energy from its initial value.

Since the integration time is limited to $10^6t_c$ from the constraint
of computer's ability, the transition of WB from the water-bag to the
isothermal distribution, which is expected to occur at later than
$10^6t_c$, is not observed. The behaviour of WB for $t\leq10^6t_c$ is
quite similar to that for $N=128$. Thus water-bag distribution is one of
quasi-equilibrium and the relaxation time to the quasi-equilibrium is
about several $10^3t_c$.

Increasing the number of the particles for IT strengthens the property
that disappearance of time-correlation which is found by the bend of the
PSD occurs the later for the larger $N$. In fact, the frequency at which
the feature of PSD changes from flat to inclined is $10^{-5}t_c^{-1}$
for $N=32$ and $10^{-6}t_c^{-1}$ for $N=128$. Together with the fact
that $\Delta(t)$ does not decrease as $t^{-1/2}$, we found that the
relaxation time, at which the trajectory covers almost whole ergodic
region, becomes longer for larger $N$.

\section{Conclusions and discussions}
\label{sec:conclusion}
We performed $N$-body simulations of the evolution of one-dimensional
self-gravitating systems for different $N$. We examined two
courses of evolution, IT and WB, which have the initial distributions of
the isothermal and water-bag, respectively. We obtained the following
lines of conclusion.
\begin{enumerate}
\item These systems relax to the isothermal distribution. The time at
which the difference between IT and WB disappears is about $10^6t_c$ for
$N=32$, $10^6t_c$ for $N=128$, and longer than $10^6t_c$ for $N=512$.

\item As $N$ increases, the water-bag distribution becomes a
quasi-equilibrium, which yields ergodic properties apparently. The
microscopic relaxation time of the water-bag, at which the time
correlation disappears, is $10^3t_c$ for $N=128$ and $10^4t_c$ for
$N=512$. The water-bag distribution lasts stably for a period orders of
magnitude longer time than the time that the correlation disappears.

\item Since the isothermal distribution is the micro-canonical
equilibrium, IT shows no evolution macroscopically. As one can see in
Figures of PSD, the microscopic relaxation time is, however, much longer
than that of the water-bag quasi-equilibrium. For $N=32$, the
microscopic relaxation time is about $10^5t_c$ and $10^6t_c$ for
$N=128$. For $N=512$, we did not observe the microscopic relaxation in
$10^6t_c$. The transition of the water-bag to the isothermal
distribution occurs in the same time scale.

\item Convergence of the maximum time-dependent Lyapunov exponents to
the different values for IT and WB indicates that the phase space is
separated into at least two ergodic-like regions which give the
quasi-equilibria.
\end{enumerate}

In the limit of $N\rightarrow\infty$ the evolution of the system, both
the isothermal and water-bag distribution are the stationary solution of
the collisionless Boltzmann-Poisson equations. The distribution function
which depends only on the individual particle energy is always the
stationary solution of the collisionless Boltzmann equation. Among these
solutions many satisfy the Poisson equation, so many class of analytic
stationary solutions for the collisionless Boltzmann-Poisson equations
exist. These solutions are expected to be also quasi-equilibria, and
they should depend on the initial condition. In fact, we have examined
another initial condition, in which all particles have the same energy.
The distribution of particles is shown in Fig.~\ref{fig:isoe}. Since
this distribution is unstable, the ellipse in the phase space is broken
rapidly, then the system approaches not to the isothermal distribution
but to the water-bag distribution. So in this case the final results are
similar to the one for WB mentioned before. Furthermore we are presently
studying the dependence of the initial condition in more details. The
result will be described in the succeeding paper.

{}From these facts, it is obvious that the phase space is not homogeneous
but contains many substructures which induce the quasi-equilibria.
Though the origin of the substructures is not clear, we have a
speculation that tori play an important role. This idea conflicts with
the usual idea that chaos is strengthen as the number of degrees of
freedom increases, thus the system approaches to ergodic. The latter
idea is supported by some numerical results, in which the volume of the
KAM tori is reduced as the number of degrees of freedom increases
\cite{gal83,kon89}. However, we believe our idea, because of the
following reasons: in the limit of $N\rightarrow \infty$, the
distribution function which depends only on the individual particle
energy is alway the stationary solution of the collisionless Boltzmann
equation. These solutions form tori because all phase elements in the
phase space (i.e. all degrees of freedom) conserves their energy, and
the motion which has the same number of integrals of motion as the
degree of freedom is a torus. Thus in the limit the system contains many
tori, which seem to produce the multi-ergodicity \cite{aiz89b}. The
other reason is that, in fact, chaos is getting weaker as N increases.
Figure \ref{fig:lyap} shows the converged value of the maximum
time-dependent Lyapunov exponents of IT and WB for various $N$. That of
IT decreases as $N^{-1/5}$ and of WB as $N^{-1/4}$.  This fact suggests
that the system approaches to near-integrable system as $N$ increases.

The property that chaos becomes weaker as the number of degrees of
freedom of the system increases seems due to the long range force,
because the force acting on a particle is the summation of
the force of all the others, thus the effect of close encounter, which
is usually believed to be the origin of chaos, becomes weaker as the
system population increases. The results for $N\leq10$ that the
stochasticity of the systems are strengthened \cite{fro75,ben79,wri84}
seem to be due to the small number of particles in the system. In such
systems, the properties of the long range force are not prominent. Thus
it is inferred that there exists the number of degrees of freedom which
maximizes the stochasticity of the system between 10 and 30.

In the three-dimensional system, the properties described above are
expected to hold, because the force is also long range. In fact,
elliptical galaxies have the stationary state which is triaxial in shape
and anisotropic in the velocity dispersion. Also the stationary states
depend on the initial conditions\cite{dez91,fun92a}. This state is
completely different from what Lynden-Bell predicted by using ergodic
hypotheses \cite{lyn67}.  This state can be explained as the
quasi-equilibrium mentioned in this paper.

\acknowledgments
It is a pleasure to thank Y. Aizawa and S. Takesue for many stimulating
discusions, and to thank K. Kitahara for introducing us the new
numerical method. A part of computations was carried out on DEC station
3000AXP of Cosmic-Ray group, Department of Physics, Kyoto University.


\begin{figure}
\caption{The water-bag initial distribution in $\mu$-space. \hfill}
\label{fig:WB}
\end{figure}

\begin{figure}
\caption{The isothermal initial distribution in $\mu$-space. \hfill}
\label{fig:IT}
\end{figure}

\begin{figure}
\caption{The cumulative specific energy distribution of the
isothermal (solid line) and water-bag (dashed line) distribution for
$N=512$. The dotted line indicates the energy at equipartition.}
\label{fig:edist}
\end{figure}

\begin{figure}
\caption{Evolution of the deviation from the equipartition, $\Delta(t)$
(bold lines), and the maximum Lyapunov exponent, $\lambda(t)$ (light
lines) for $N=32$. The solid and dashed lines represents IT and WB,
respectively. The unit of time is $t_c$.}
\label{fig:bi5a}
\end{figure}

\begin{figure}
\caption{Power spectrum densities of IT (solid line) and WB (dashed
line) for $N=32$. The unit of frequency is $t_c^{-1}$.}
\label{fig:bi5b}
\end{figure}

\begin{figure}
\caption{Same as Fig. 4, but for $N=128$.}
\label{fig:bi7a}
\end{figure}

\begin{figure}
\caption{Same as Fig. 5, but for $N=128$.}
\label{fig:bi7b}
\end{figure}

\begin{figure}
\caption{Cumulative energy distributions, $\hat{\nu}(\varepsilon)$, of
IT at the beginning, $t=0$, (solid line), WB at the beginning (dashed
line), and WB at $t=10^6t_c$ (dotted line).}
\label{fig:bi7c}
\end{figure}

\begin{figure}
\caption{Cumulative energy distributions, $\hat{\nu}(\varepsilon)$, of
IT (solid line) and WB (dashed line) at $t=10^7t_c$.}
\label{fig:bi7d}
\end{figure}

\begin{figure}
\caption{Same as Fig. 4, but for $N=512$.}
\label{fig:bi9a}
\end{figure}

\begin{figure}
\caption{Same as Fig. 5, but for $N=512$.}
\label{fig:bi9b}
\end{figure}

\begin{figure}
\caption{Distribution of particle in a stationary solution of
collisionless Boltzmann-Poisson equations. All the particles have the
same energy.}
\label{fig:isoe}
\end{figure}

\begin{figure}
\caption{The maximum Lyapunov exponents of IT (solid line) and WB
(dashed line) after convergence.}
\label{fig:lyap}
\end{figure}

\end{document}